\documentclass[journal]{IEEEtran}
\usepackage{ifpdf}

\usepackage{wrapfig}

\usepackage{cite}

\ifCLASSINFOpdf
\usepackage[pdftex]{graphicx}
\else
\fi
\usepackage{amsmath}
\usepackage{array}
\usepackage{url}

\usepackage{amsmath,amssymb,amsfonts}
\usepackage{graphicx}
\usepackage{textcomp}
\usepackage{wrapfig}

\usepackage{url}
\usepackage{color}
\usepackage{stfloats}
\usepackage{subcaption}
\usepackage{multirow}
\usepackage{algorithm}
\usepackage[noend]{algpseudocode}
\usepackage{gensymb}

\begin{document}
	%
	%
	%
	%
	
	\title{DepthNet: Real-Time LiDAR Point Cloud Depth Completion for Autonomous Vehicles}
	\author{Lin~Bai,~\IEEEmembership{Student~Member,~IEEE,}
			Yiming~Zhao,~\IEEEmembership{Student~Member,~IEEE,}
			Mahdi Elhousni,
			and~Xinming~Huang*,~\IEEEmembership{Senior~Member,~IEEE}
	\thanks{L. Bai, Y. Zhao, M. Elhousni and X. Huang are with the Department
		of Electrical and Computer Engineering, Worcester Polytechnic Institute, Worcester,
		MA, 01609 USA (e-mail: {\{lbai2,yzhao7,melhousni,xhuang\}@wpi.edu}).}}
	
	%
	%

	\markboth{Journal of \LaTeX\ Class Files,~Vol.~14, No.~8, August~2015}%
	{Shell \MakeLowercase{\textit{et al.}}: Bare Demo of IEEEtran.cls for IEEE Journals}
	%



	\maketitle
	
\begin{abstract}
Autonomous vehicles rely heavily on sensors such as camera and LiDAR, which provide real-time information about their surroundings for the tasks of perception, planning and control. Typically a LiDAR can only provide sparse point cloud owing to a limited number of scanning lines. By employing depth completion, a dense depth map can be generated by assigning each camera pixel a corresponding depth value. However, the existing depth completion convolutional neural networks are very complex that requires high-end GPUs for processing, and thus they are not applicable to real-time autonomous driving. In this paper, a light-weight network is proposed for the task of LiDAR point cloud depth completion. With an astonishing 96.2\% reduction in the number of parameters, it still achieves comparable performance (9.3\% better in MAE but 3.9\% worse in RMSE) to the state-of-the-art network. For real-time embedded platforms, depthwise separable technique is applied to both convolution and deconvolution operations and the number of parameters decreases further by a factor of 7.3, with only a small percentage increase in RMSE and MAE performance. Moreover, a system-on-chip architecture for depth completion is developed on a PYNQ-based FPGA platform that achieves real-time processing for HDL-64E LiDAR at the speed \textcolor{black}{11.1} frame per second.
\end{abstract}

\begin{IEEEkeywords}
LiDAR, point cloud, depth completion, convolutional neural network, FPGA
\end{IEEEkeywords}

\IEEEpeerreviewmaketitle

\section{Introduction}
\IEEEPARstart{I}{n} recent years, autonomous vehicles have become a rapidly evolving technology that may revolutionize mobility and transportation systems. To accurately sense vehicle surroundings,
cameras are often employed to provide a 2D description of the space. However, in order to transition into the 3D space, two options are usually employed, RGB-D cameras or LiDARs. Nevertheless, limited by the short range (around 10 meters) and weak energy, RGB-D cameras are mostly suited for indoor applications. On the other hand, modern LiDARs are capable of supplying accurate distance information up to 100 meters. In addition, LiDAR performance does not depend on the changes in lighting conditions. These advantages make LiDAR an ideal 3D sensor for outdoor applications such as autonomous vehicles. 

One drawback of LiDAR sensor is its data sparsity: When mapping a Velodyne 64 line LiDAR HDL-64E point cloud to its corresponding high-resolution image obtained from a camera, only about 10\% of the pixels have depth values. Especially, when laser scan lines encounter transparent or reflective surfaces such as car windows, the depth values are void. Therefore, depth completion is an important task that is aimed to data sparsity problem by generating a complete depth map for every pixels in the camera image and also making corrections of some void values. It is somewhat similar to an interpolation process, but we have to consider the feature of the objects in the 3D point cloud. Depth completion results a dense and precise depth map. When combined with RGB images, depth completion makes it possible for an autonomous vehicle to detect objects in 3D space and predict their movement accurately.

Most of the existing depth completion methods are based on convolutional neural networks (CNNs) that are very complex and can only run on high-power GPUs, such as Nvidia GTX 2080Ti, TITAN X, etc. Considering the limited power supply available in an autonomous vehicle, CNNs targeted on a real-time embedded platforms are much desirable. This calls for a novel CNN architecture with orders of magnitude reduction in the number of parameters and operations while maintaining a comparable performance. In this paper, we focus on addressing this important issue by introducing a two-stage learning method (coarse estimation stage and residual learning stage) and depthwise separable technique. Furthermore, we target the proposed efficient CNN architecture on a PYNQ-based MPSoC FPGA platform and demonstrate real-time LiDAR point cloud processing with superior operation-power ratio.

The contributions of this work can be summarized as follows:
\begin{enumerate}
    \item A light-weight depth completion neural network is proposed with residual learning method. This neural network significantly reduces the number of parameters by a factor of \textcolor{black}{26.1}, while achieves comparable error performance. Further optimization with depthwise separable technique, the number of parameters is decreased to only 0.53\% of the state-of-the-art (SOTA) network \cite{ma2019self}. The error performance result is comparable to that of SOTA when evaluating with the depth completion dataset.  
    \item An efficient hardware architecture is designed for the LiDAR depth completion network. In particular, deconvolution operations are implemented by avoiding all extra multiplication with zeros. By carefully balancing on-chip memory and multipliers, the FPGA implementation can execute the proposed depth completion neural network in real-time at 11.1 frames per second (fps).
    \item A PYNQ-based LiDAR sensing and processing system is introduced. By migrating the Velodyne LiDAR driver to Linux system, one can use Python command to receive point cloud data and execute the depth completion neural network on hardware. An example work for VLP-16 LiDAR is open sourced and made available at \url{https://github.com/linbaiwpi/VLP16_driver_on_PYNQ}.
\end{enumerate}

\section{Related Work}
In general, there are two classes of methods for the depth completion problem, namely classical methods and learning-based methods. The former ones utilize the traditional computer vision algorithms to complete the depth map. Unfortunately, most of them are applicable to RGB-D camera, which are not suitable for LiDAR point cloud. Ku \cite{ku2018defense} addressed the point cloud depth completion issue by using only basic computer vision operations such as dilation, smoothing, etc. This solution achieved comparable performance to some learning based methods even.

The other solution, learning-based methods, dominant the depth completion solutions, due to the huge success of deep learning for computer vision tasks. Chodosh et al. provided a solution for depth completion by combining compressed sensing and deep learning on Alternating Direction Neural Network (ADNN) framework \cite{chodosh2018deep}. Eldesokey et al. further designed a network composed by normalized convolution layer which only contains two channels, depth map and confidential map \cite{eldesokey2018propagating}. In contrast to specially designed convolution kernel above, Ma et al. \cite{ma2019self} solved this problem by directly putting the raw sparse depth image into a large 34-layer network. This SOTA performance indicated the regular 2D convolution is able to solve the sparsity if the network is deep enough. In addition, self-supervised learning methods were adopted recently aiming to avoiding the heavy manually labeling work. Ma further extended a self-supervised depth prediction framework to depth completion by feeding the sparse points into the network and treat them as the ground truth for corresponding pixels.

Because CNNs usually require huge computation capability which results in very high power consumption, tremendous research efforts have been dedicated to high-performance and low-power CNN accelerators for embedded devices such as FPGAs. In \cite{zhang2015optimizing}, the authors proposed a novel architecture for process element array. By exploring the design space, this design well balanced the computation capability and bandwidth requirement. Some works also focused on deconvolution. Liu et al. proposed one CNN architecture for segmentation, where convolution and deconvolution as two peripherals loaded on the system bus \cite{liu2018optimizing}. In \cite{zhang2017design}, a high performance deconvolution module was proposed, in which reversed looping and stride hold skipping were employed to improve the performance. Some other implementations were targeted toward specific applications like autonomous driving \cite{lyu2018real}\cite{peng2019multi}.

\section{Proposed Neural Network}

\begin{figure*}[htbp]
    \centering
    \includegraphics[width=\linewidth]{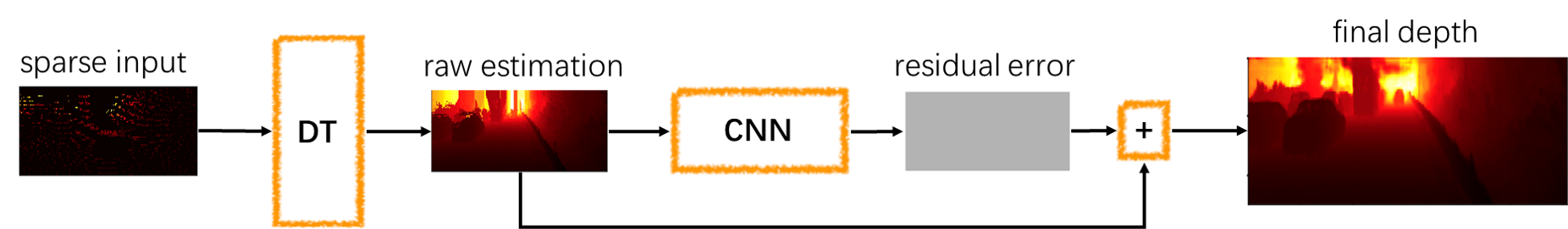}
    \caption{The deep learning framework for depth completion from LiDAR, where DT is distance transform and CNN is the neural network for residual error learning}
    \label{fig:cnn_frame}
\end{figure*}

Inspired by the residual learning \cite{he2016deep} and the traditional computer vision method \cite{ku2018defense}, we propose a two-stage solution to complete the sparse depth map. In the first stage, coarse depth information is predicted using distance transformation algorithm. And this raw result is further refined by a residual learning network. The entire deep learning framework is shown in Fig.~\ref{fig:cnn_frame}.

\subsection{Network Structure}
In \cite{ku2018defense}, the empty pixel is filled by dilation operation in very small local patches, which results in a reasonable estimation. This means the depth values of neighbouring pixels are probably very close to each other. Inspired by this finding, a nearest neighbor operation is applied to get a raw depth map before feeding it into the CNN, as a raw estimation of empty pixels. Distance transform, which has a linear complexity, is adopted as the method to find the nearest neighbor. Considering the accuracy of depth completion, a CNN is introduced to refine the raw estimation. Mathematically that is to find the residual error. The final depth map is the combination of raw estimation and residual error.  The raw sparse depth map from LiDAR is the input. And then a raw estimation is generated by distance transform for CNN, who refines the raw estimation by finding the residual error. In the end, the final depth map is the raw estimation corrected by the predicted residual error.

\begin{figure}[htbp]
     \centering
     \begin{subfigure}[b]{0.45\columnwidth}
         \includegraphics[width=\textwidth]{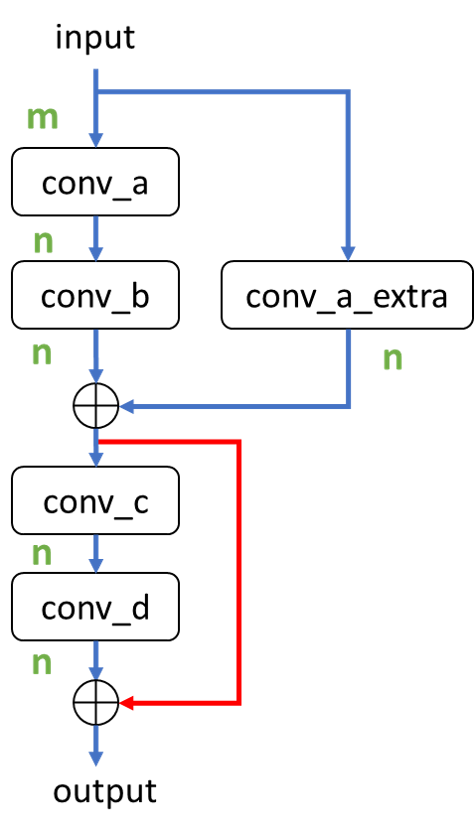}
         \caption{}
         \label{fig:encoder}
     \end{subfigure}
     \hfill
     \begin{subfigure}[b]{0.45\columnwidth}
         \includegraphics[width=\textwidth]{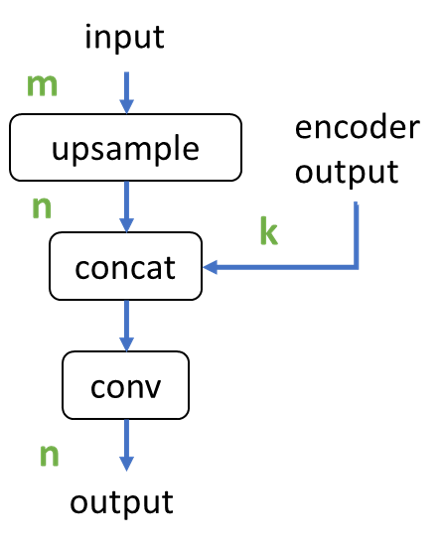}
         \caption{}
         \label{fig:decoder}
     \end{subfigure}
     \caption{(a) is the structure of encoder block (block 4 has no feed forward marked in red) and (b) is the structure of decoder block}
     \label{fig:cnn_block_arch}
\end{figure}

\begin{figure}[htbp]
    \centering
    \includegraphics[width=0.9\columnwidth]{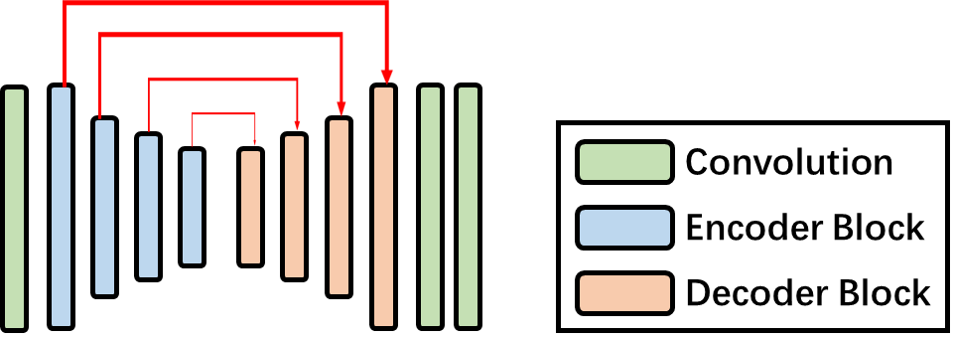}
    \caption{The structure of proposed CNN}
    \label{fig:struct}
\end{figure}

The neural network structure is illustrated in Fig.~\ref{fig:struct} and detailed  information (input and output size, channel depth) is listed in Tab.~\ref{tab:cnn_arch}. It is basically an encoder-decoder architecture. The encoder block structures is demonstrated in Fig.~\ref{fig:encoder}. Except the kernel size of 'conv\_a\_extra' is $1\times 1$, the kernel size of the rest convolutions are $3\times 3$. For encoder block 4, there is no feed forward line marked in red. Fig.~\ref{fig:decoder} illustrates the structure of decoder blocks, in which, the 'upsample' operation is deconvolution. Instead of using max pooling, all the down sampling in this network is done by convolution with strides equal to 2 in 'conv\_a' and 'conv\_a\_extra'. The number of channels for each layer is indicated by 'Channel' column in Tab.~\ref{tab:cnn_arch}.

\subsection{Network Optimization}
Different from the existing neural network based methods \cite{chodosh2018deep}\cite{eldesokey2018propagating}\cite{uhrig2017sparsity}, the results from \cite{ma2019self} concluded that a very deep regular convolution network can also complete depth maps very well. A deep ResNet architecture with 34 layers is utilized in their solution. However, the computation complexity of training ResNet-34 for high-resolution input images is too high to fit on a GeForce RTX2080 Ti GPU with 11GB graphic memory. Therefore, a much lighter structure similar to ResNet-18 is adopted in the encoder of our proposed CNN. By balancing the computation complexity and performance, ResNet-18 is easier to train and more suitable for mobile devices.

\begin{table}[htbp]
\centering
\caption{Structure of the proposed CNN (E = Encoder, D = Decoder)}
    \begin{tabular}{|l|l|l|l|} 
        \hline
        Name & Input Size & Output Size & Channel(m,n,k) \\
        \hline
        In conv   & $1216\times 256\times 1$ & $1216\times 256\times 32$ & \\
        \hline
        E block 1 & $1216\times 256\times 32$ & $1216\times 256\times 32$ & \textcolor{black}{32, 32, 0} \\
        \hline
        E block 2 & $1216\times 256\times 32$ & $608\times 128\times 32$ & \textcolor{black}{32, 32, 0} \\
        \hline
        E block 3 & $608\times 128\times 32$ & $304\times 64\times 64$ & \textcolor{black}{32, 64, 0} \\
        \hline
        E block 4 & $304\times 64\times 64$ & $152\times 32\times 128$ & \textcolor{black}{64, 128, 0} \\
        \hline
        D block 1 & $152\times 32\times 128$ & $304\times 64\times 64$ & \textcolor{black}{128, 64, 64} \\
        \hline
        D block 2 & $304\times 64\times 64$ & $608\times 128\times 32$ & \textcolor{black}{64, 32, 32} \\
        \hline
        D block 3 & $608\times 128\times 32$ & $1216\times 256\times 32$ & \textcolor{black}{32, 32, 32} \\
        \hline
        Out conv1 & $1216\times 256\times 32$ & $1216\times 256\times 32$ & \\
        \hline
        Out conv2 & $1216\times 256\times 32$ & $1216\times 256\times 1$ & \\
        \hline
    \end{tabular}
\label{tab:cnn_arch}
\end{table}

To further squeeze the proposed network, we applied the depthwise separable concept to both convolution and deconvolution. Depthwise separable convolution operation is exactly the same as the one used in MobilNetV2\cite{sandler2018mobilenetv2}. Similarly, for depthwise separable deconvolution, input feature map has been convoluted for each channel separately. Subsequently, a pointwise convolution is applied to generate the output feature map.

\subsection{Training}
The input feature map size is $256\times 1216\times 1$, which is generated by following the methods mentioned in \cite{van2019sparse}\cite{tang2019learning}. As the output of CNN, residual error is normalized to $[0,~1]$ as described in \cite{ma2019self}. The empty region of output image is filled with the value on top of each line, like \cite{ku2018defense}.

The proposed CNN is implemented by TensorFlow. During training, the batch size is set to 4, and number of epochs is 6. The initial learning rate is $10^{-4}$ with decreasing by a factor of 2 after each epoch. The CNN is trained using the loss of Mean Square Errors (MSE) and the Adam optimizer function. The training dataset is from KITTI dataset, depth completion task, which supplies collection of sparse LiDAR scans and corresponding semi-dense depth maps as ground truth. Totally, there are 85,898 scans as training samples, 1,000 scans as validation samples and 1,000 scans as test samples.

\subsection{Evaluation}

\begin{figure*}[t]
    \centering
    \includegraphics[height=100mm,width=\linewidth]{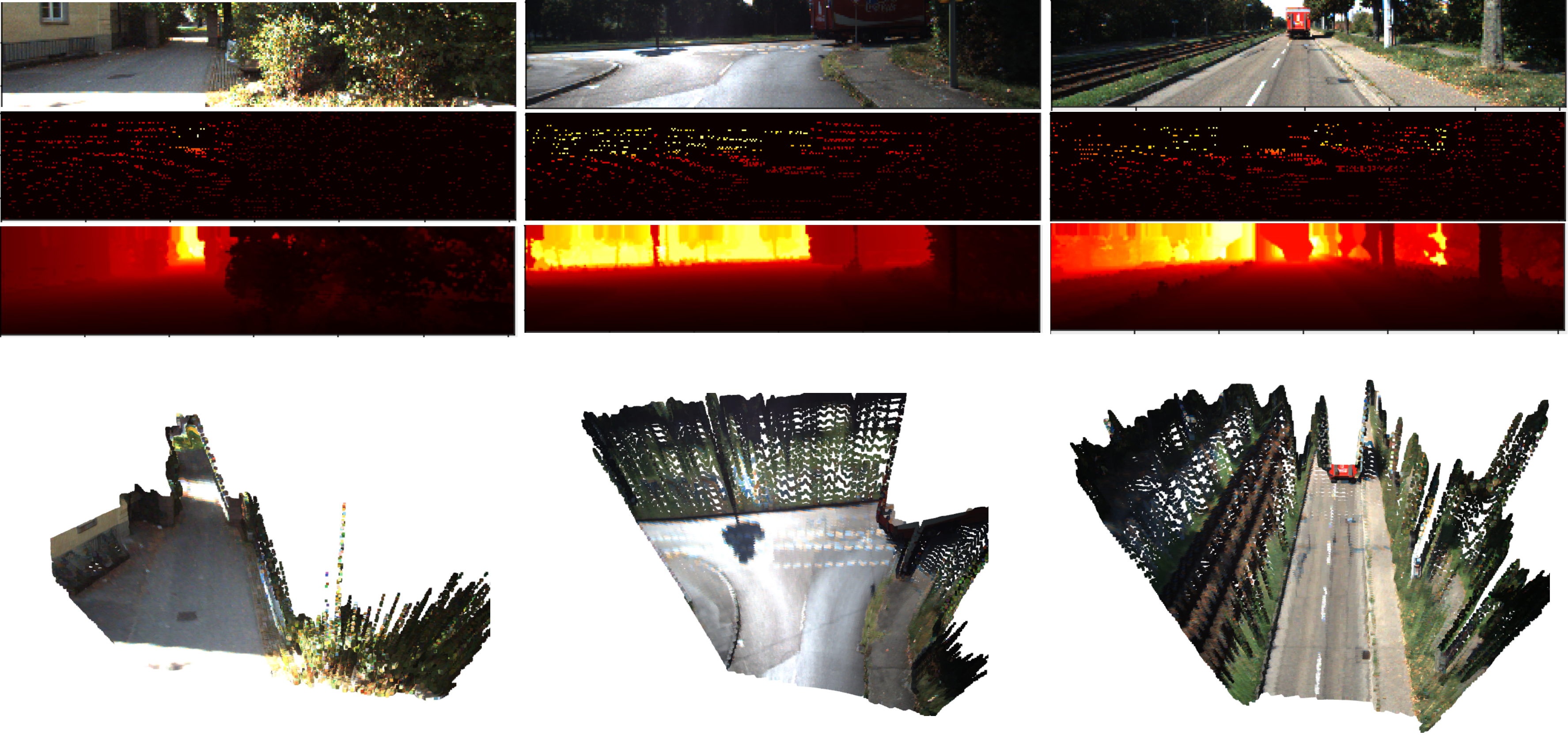}
    \caption{Illustration of the proposed depth completion framework performance. From top to bottom: 1) RGB image, 2) point cloud projected to image coordinate, 3) output of DepthNet using point cloud only, 4) 3D depth vision rebuilding}
    \label{fig:cnn_result}
\end{figure*}

In Fig.~\ref{fig:cnn_result}, the performance of our proposed depth completion framework is illustrated. In the upper part of the figure, RGB image, sparse depth map projected from point cloud and the predicted dense depth map are pasted from top to bottom respectively. Based on RGB images and their corresponding predicted depth maps, the 3D RGB model is built and shown at the bottom of Fig.~\ref{fig:cnn_result}. The images and point clouds are all from KITTI depth completion dataset.

The predicted results are evaluated by the following 4 metrics:
\begin{enumerate}
    \item RMSE: Root Mean Squared Error [mm]
    \item MAE: Mean Absolute Error [mm]
    \item iRMSE: RMSE of the inverse depth [1/km]
    \item iMAE: MAE of the inverse depth [1/km]
\end{enumerate}

The performance comparison between our proposed networks and the SOTA networks are given in Tab.~\ref{tab:peform_cmp}. Our network (without depthwise separable operation) is 9.3\% better and 3.9\% worse than the SOTA result \cite{ma2019self}, in terms of MAE and RMSE respectively. Considering their properties, the MAE represents the average error of the predicted depth map. 
$$MAE = \frac{1}{n}\sum{|y_{predict}-y_{true}|}$$

While the RMSE is more sensitive to the pixels with large errors in the predicted depth map.
$$RMSE = \sqrt{\frac{1}{n}\sum{(y_{predict}-y_{true})^2}}$$
Lower MAE but higher RMSE means that our network has less average error than that of \cite{ma2019self}, but has more pixels with large error. Besides, the number of parameters of our network is $9.73\times10^5$, which is only 3.8\% of the SOTA network $2.54\times10^7$ \cite{ma2019self}.

To further squeeze the network for embedded platforms, we applies the depthwise separable (DS) technique to both convolutions and deconvolutions. This results in 12.8\% higher in terms of RMSE and 2.3\% less in terms of MAE. From the parameters number point of view, our network with DS requires $1.34\times10^5$ parameters. Comparing that of SOTA network, our network with DS reduces the number of parameters by a factor of 189.56. What's more, the number of operations decreases accordingly. These reductions make this network more suitable for embedded platform in terms of computation complexity and bandwidth requirement.

\begin{table}[htbp]
\centering
\caption{Performance comparison of LiDAR only depth completion CNNs (DS = depthwise separable)}
    \begin{tabular}{|c|c|c|c|c|} 
        \hline
        \multirow{2}{4em}{Network} & iRMSE & iMAE & RMSE & MAE \\
             & (1/km) & (1/km) & (mm) & (mm)\\
        \hline
        \textbf{Ours(without DS)} & \textbf{2.99} & \textbf{1.09} & 991.88 & \textbf{261.67} \\ 
        \hline
        Sparse-to-Dense\cite{ma2019self} & 3.21 & 1.35 & \textbf{954.36} & 288.64 \\
        \hline
        \textbf{Ours(with DS)} & 3.43 & 1.21 & 1077.22 & 282.02\\ 
        \hline
        NConv-CNN\cite{eldesokey2018propagating} & 4.67 & 1.52 & 1268.22 & 360.28 \\
        \hline
        ADNN\cite{chodosh2018deep} & 59.39 & 3.19 & 1325.37 & 439.48 \\
        \hline
        SparseConvs\cite{uhrig2017sparsity} & 4.94 & 1.78 & 1601.33 & 481.27 \\
        \hline
    \end{tabular}
\label{tab:peform_cmp}
\end{table}

\section{Optimization for Hardware}\label{sec:opt}

\subsection{Network Optimization}\label{sec:net_opt}
\begin{figure}[!]
    \centering
    \begin{subfigure}[b]{0.9\linewidth}
        \centering
        \includegraphics[width=\textwidth]{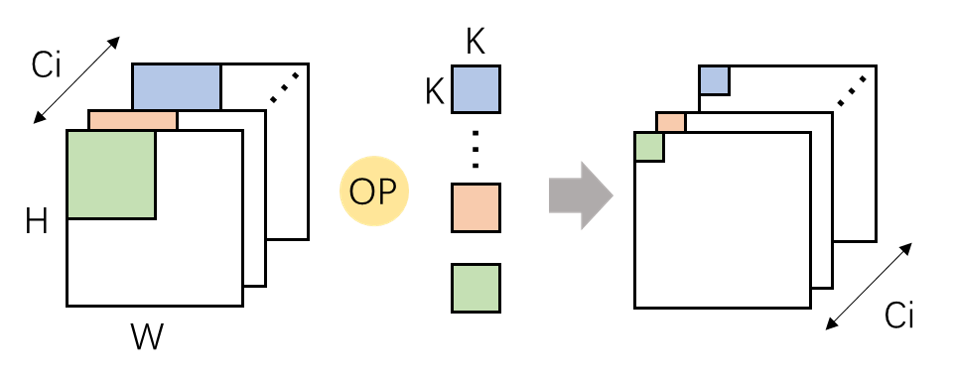}
        \caption{Step 1: depthwise operation, for DSC, OP is convolution, and for DSD, OP is deconvolution}
        \label{fig:depth_op}
    \end{subfigure}
    \newline
    \begin{subfigure}[b]{0.9\linewidth}
        \centering
        \includegraphics[width=\textwidth]{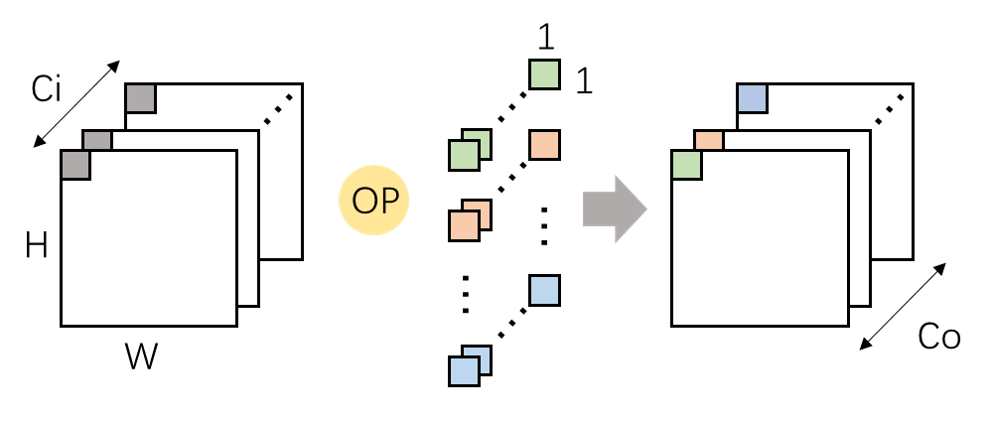}
        \caption{Step 2: pointwise operation, for both DSC and DSD, OP is convolution}
        \label{fig:point_op}
    \end{subfigure}
    \caption{Operation comparison between depthwise separable convolution (DSC) and depthwise separable deconvolution (DSD)}
    \label{fig:ds_op}
\end{figure}

To further squeeze the proposed network, we applied the depthwise separable concept to both convolution and deconvolution. Depthwise separable convolution operation is exactly the same as the one used in MobilNetV2\cite{sandler2018mobilenetv2}. Similarly, depthwise separable deconvolution completes the operation in two step as well. The input feature maps have been deconvoluted for each channel separately (depthwise operation). Subsequently, a pointwise convolution is applied to generate the output feature map (pointwise operation). Fig.~\ref{fig:ds_op} illustrates the procedures for depthwise separable convolution and deconvolution. During depthwise operation in Fig.~\ref{fig:depth_op}, the operations are not the same for convolution and deconvolution. Besides, the output feature map size doubles the input size in deconvolution, while they are the same in convolution. Both convolution and deconvolution share the same pointwise operation.

When comparing with their standard counterpart, the parameters number for both convolution and deconvolution consume only $\frac{1}{C_o}+\frac{1}{K^2}$.

$$\frac{without DS}{with DS} = \frac{K\cdot K\cdot C_i + 1\cdot 1\cdot C_i\cdot C_o}{K\cdot K\cdot C_i\cdot C_o} = \frac{1}{C_o}+\frac{1}{K^2}$$

\subsection{Loop Optimization}\label{sec:loop_opt}
Since all standard convolutions are replaced by depthwise separable convolution. The ordinary loop optimization are down-graded into 3 level nest loops for both depthwise convolution (Alg.~\ref{alg:loop_dw}) and pointwise convolution (Fig.~\ref{alg:loop_pw})\cite{ma2017optimizing}. Concerning to depthwise separable deconvolution, both depthwise and pointwise operations share the same loops as their convolution counterparts. The only difference is the kernel size in depthwise operation \cite{bai_iscas2020}.

\subsubsection{depthwise operation}
Recalling our previous work \cite{bai_iscas2020}, convolution and deconvolution could share the same process element if loop 1 is completely unrolled. Besides, limited by number of multipliers and BRAMs on FPGA, the input feature maps are partitioned and processed sequentially. In addition, the loop 3 is partially unrolled. Loop 2 remains intact.

\begin{algorithm}
    \caption{Loops for depthwise operation}
    \label{alg:loop_dw}
\begin{algorithmic}[htbp]
    \For{no in Nof}\Comment{channel,loop-3}
        \For{(y,x) in (Noy,Nox)}\Comment{feature map,loop-2}
            \For{(ky,kx) in (K,K)}\Comment{kernel,loop-1}
                \State $F_{out}$[no,y,x]+=
                \State $F_{in}$[no,y-ky,x-kx] *$K$[no,ky,kx]
            \EndFor
        \EndFor
    \EndFor
\end{algorithmic}
\end{algorithm}

\subsubsection{pointwise operation}
The $1\times 1$ convolution is mathematically matrix multiplication, which is a 3 cascaded loops. To share the same feature map buffers with depthwise operations, the loop 1 is partially unrolled with the same partition factor as depthwise loop 3.

\begin{algorithm}
    \caption{Loops for pointwise operation}
    \label{alg:loop_pw}
\begin{algorithmic}[htbp]
    \For{no in Nof}\Comment{output channel,loop-3}
        \For{(y,x) in (Noy,Nox)}\Comment{feature map,loop-2}
            \For{ni in Nif}\Comment{input channel,loop-1}
                \State $F_{out}$[no,y,x]+=\State $F_{in}$[ni,y,x] *$K$[no,ni]
            \EndFor
        \EndFor
    \EndFor
\end{algorithmic}
\end{algorithm}

\subsection{Deconvolution Optimization}\label{sec:opt_dc}
\begin{figure}[htbp]
	\centering
	\includegraphics[width=\columnwidth]{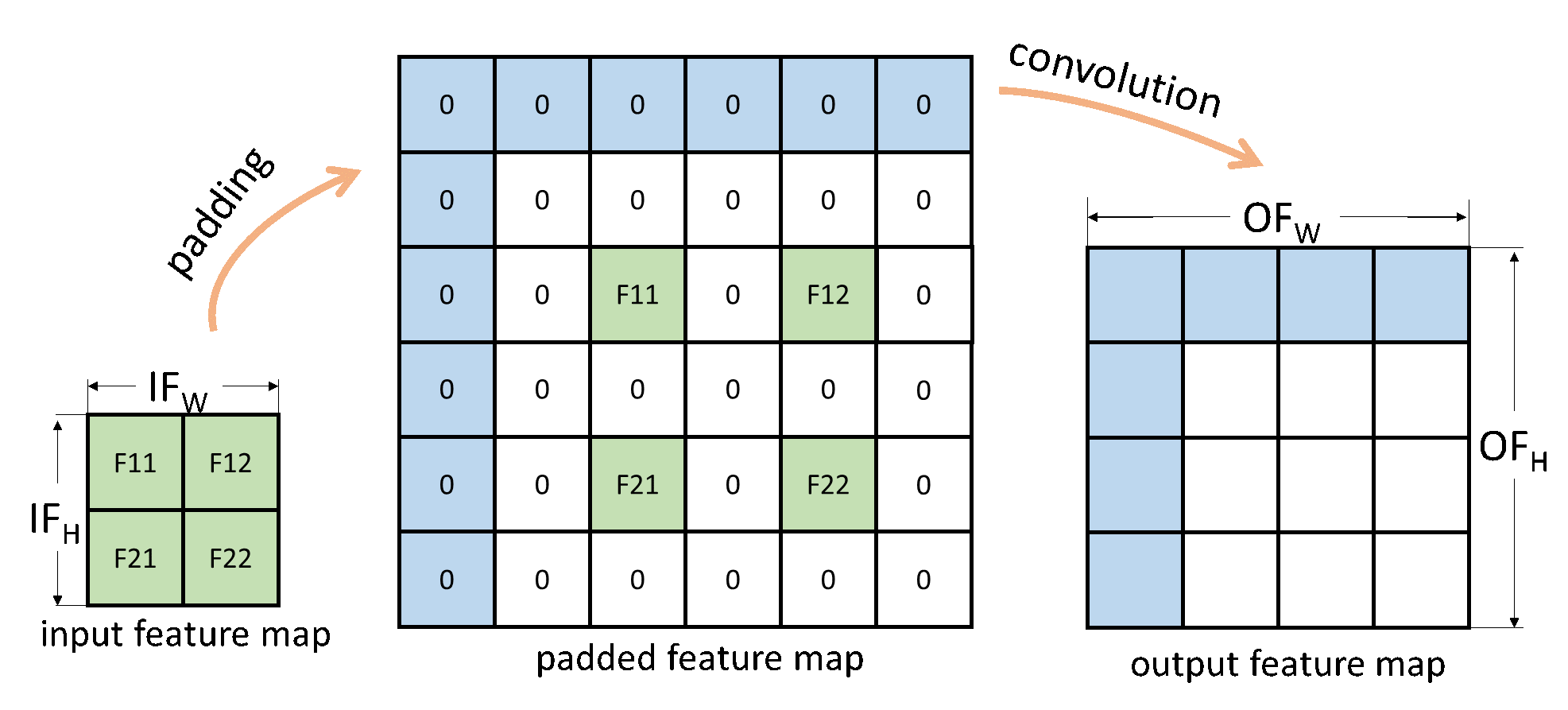}
	\caption{Naive deconvolution using padding and convolution}
	\label{fig:deconv_op}
\end{figure}

As a learnable technique for upsampling, deconvolution is widely used in depth completion tasks. A naive deconvolution operation is demonstrated in Fig.~\ref{fig:deconv_op}, where a convolution unit is reused. This naive deconvolution consists of the following two steps:
\begin{enumerate}
    \item feature map padding: the input feature map is padded from $I\!F_W\!\times\!I\!F_H$ to $(2\cdot\!I\!F_W+1)\!\times\!(2\cdot\!I\!F_H+1)$. The padded zeros are marked in blue and white in Fig.~\ref{fig:deconv_op}. The blue zeros are compulsory is twice the size is required.
    \item convolution: applying the convolution to the padded feature map.
\end{enumerate}

Based on the description above, most of the computation are wasted in multiplication by zeros. Avoiding these meaningless multiplication will boost the deconvolution speed dramatically.

\begin{figure}[htbp]
	\centering
	\includegraphics[width=0.9\columnwidth]{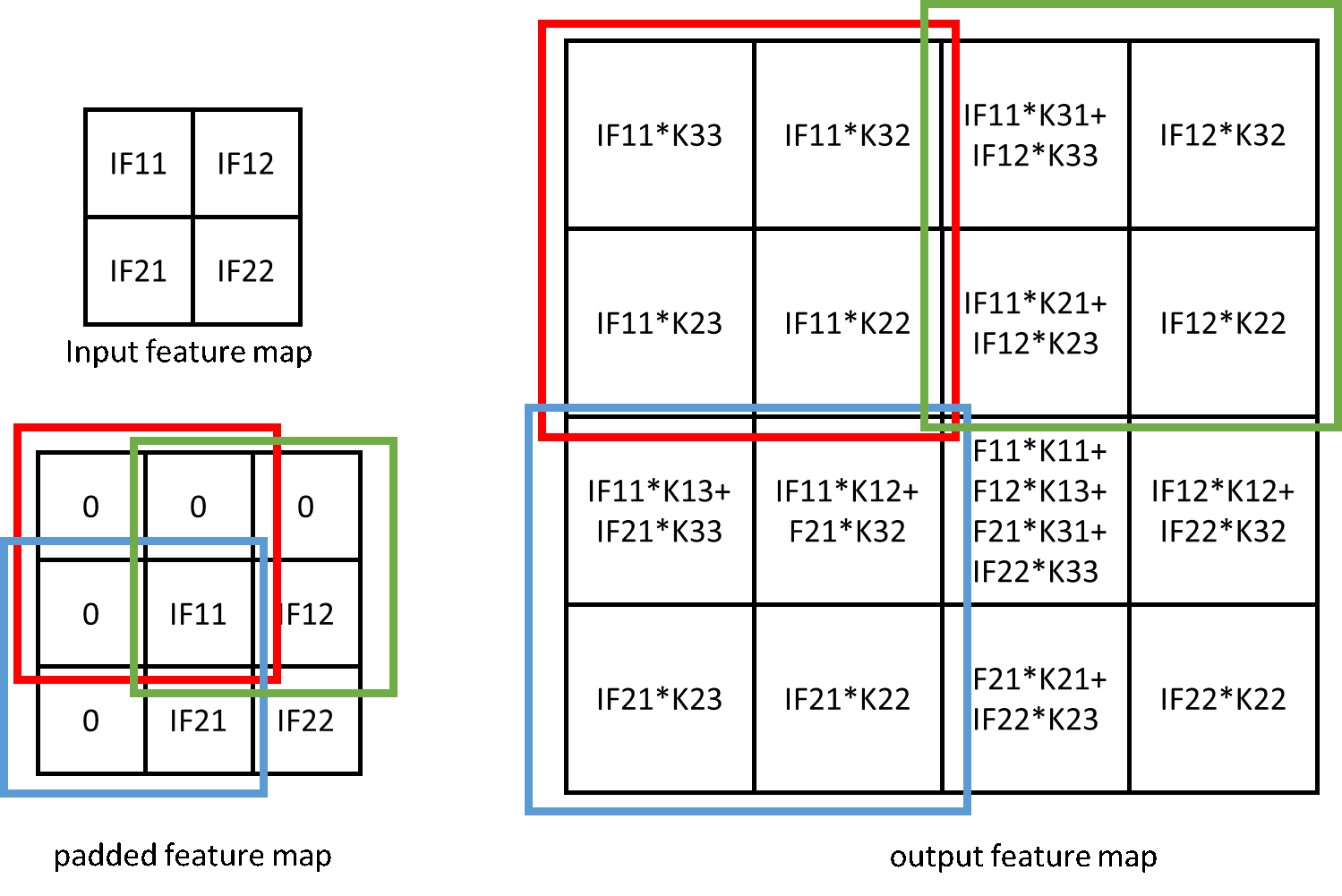}
	\caption[Caption for LOF]{Optimization of deconvolution\footnotemark}
	\label{fig:deconv_result}
\end{figure}

An efficient deconvolution method is utilized in this paper. Concerning to the same task in Fig.~\ref{fig:deconv_op}, the equations are presented in (\ref{eq:deconv_1})-(\ref{eq:deconv_4}) accompanied by Fig.~\ref{fig:deconv_result}. The procedure is divided into three steps \cite{bai_iscas2020}:
\begin{enumerate}
    \item padding input feature map: extra top row and left column is required
    \item scanning the padded feature map by sliding window in $2\times2$
    \item applying the equations to generate output feature map by $2\times 2$ patch
\end{enumerate}

\begin{align}
    O\!F_{11}\!=&I\!F_{11}\!\cdot\!K_{11}\!+\!I\!F_{12}\!\cdot\!K_{13}\!+\!I\!F_{21}\!\cdot\!K_{31}\!+\!I\!F_{22}\!\cdot\!K_{33}\label{eq:deconv_1}\\
    O\!F_{12}\!=&I\!F_{12}\!\cdot\!K_{12}\!+\!I\!F_{22}\!\cdot\!K_{32}\\
    O\!F_{21}\!=&I\!F_{21}\!\cdot\!K_{21}\!+\!I\!F_{22}\!\cdot\!K_{23}\\
    O\!F_{22}\!=&I\!F_{22}\!\cdot\!K_{22}\label{eq:deconv_4}
\end{align}

Three examples marked in red, green and blue squares respectively are represented in Fig.~\ref{fig:deconv_result}. By using this, most multiplication by zeros are saved, only the padded $I\!F_W+\!I\!F_H+1$ zeros.

\footnotetext{In TensorFlow, deconvolutions require the kernel rotated $180\degree$ before calculating. However, there is no rotation action in this example for easier description.}

\section{System Architecture}\label{sec:hardware}
The entire deep learning framework as shown in Fig.~\ref{fig:cnn_frame} for depth completion is partitioned and assigned into ARM processor and FPGA logic. The data capturing and distance transform are handled by OpenCV running on the ARM processor, while the CNN inference part runs on the FPGA logic.

\subsection{Software Task}
\begin{figure}[hbpt]
    \centering
    \includegraphics[width=\columnwidth]{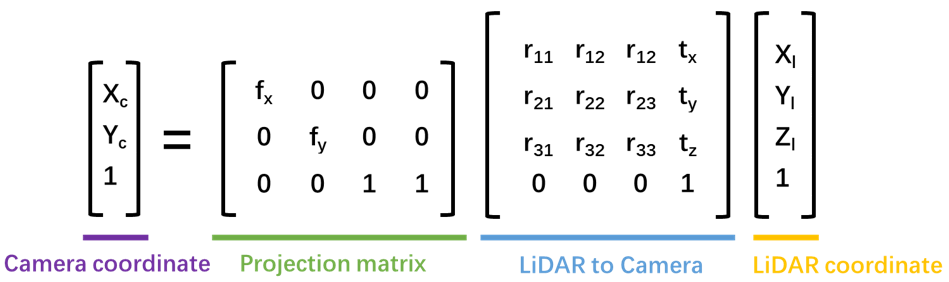}
    \caption{Sparse depth map generation, which is coordinate projection from LiDAR to camera}
    \label{fig:project}
\end{figure}

The ARM processor prepares the sparse depth map for the CNN on FPGA side. This contains three steps:
\begin{enumerate}
    \item Point cloud capturing: The Velodyne LiDAR driver is implanted as Linux dynamic library and ARM runs the driver to capture point cloud periodically.
    \item Sparse depth map generation: It projects the 3D LiDAR coordinate into 2D camera coordinate. Mathematically, this process is matrix multiplication (Fig~\ref{fig:project}).
    \item Distance transform: as mentioned before, it supplies a raw estimation for CNN.
\end{enumerate}

\subsection{Hardware Overview}
The CNN inference accelerator consists of the following three parts: 1) Process Engine (PE) computing pointwise convolution, depthwise convolution and deconvolution, and activation function LeakyReLU; 2) Buffers for weights, bias and intermediate feature maps; 3) Control logic which determines the data routing between DDR memory and buffers, between PE and buffers, etc.

\subsection{Process Engine}
\begin{figure}[hbpt]
    \centering
    \includegraphics[width=0.8\columnwidth]{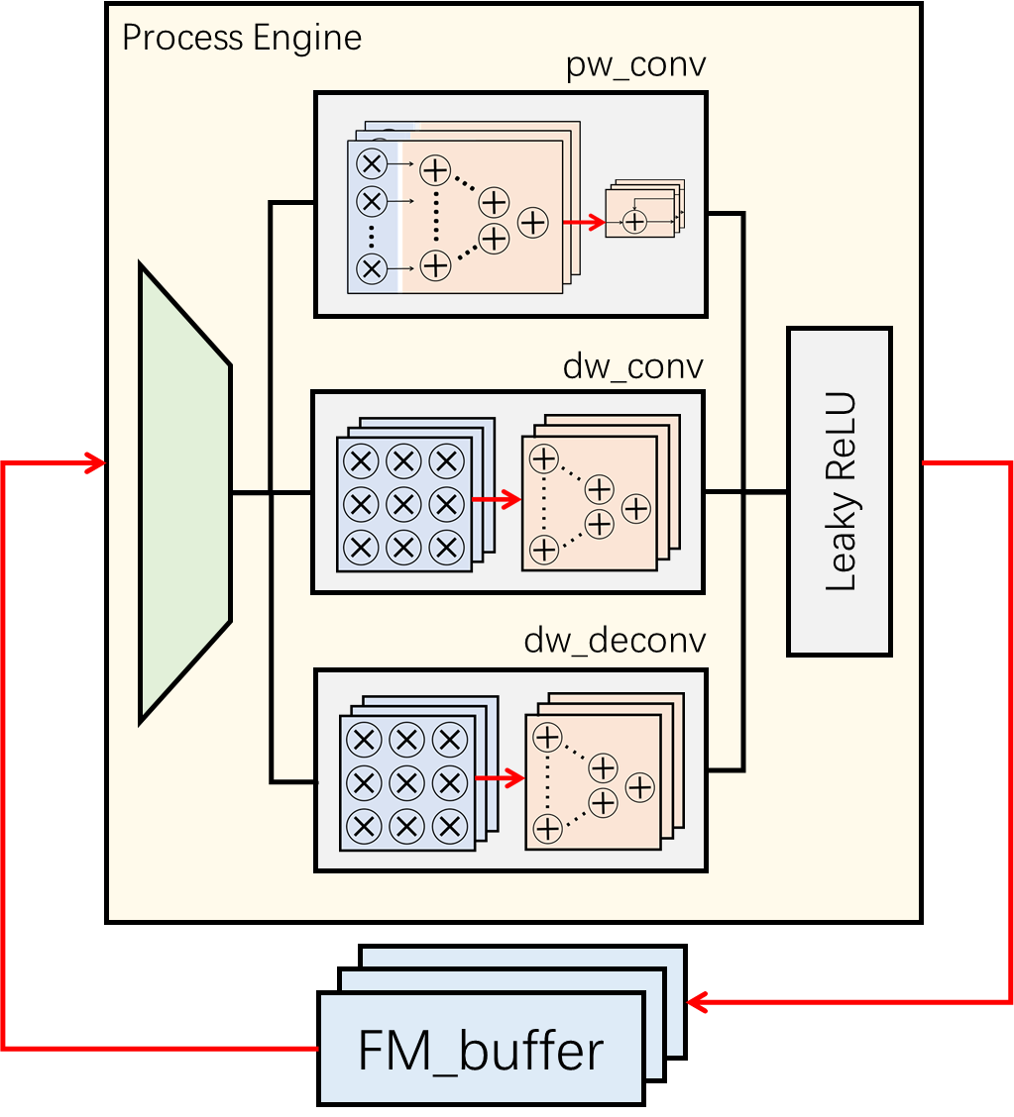}
    \caption{Structure of process engine. The upper one is pointwise convolution (pw\_conv), the middle one is depthwise convolution (dw\_conv) and the bottom one is depwthwise deconvolution (dw\_deconv).}
    \label{fig:pe_arch}
\end{figure}

According to the structure of neural network in Tab.~\ref{tab:cnn_arch} and Fig.~\ref{fig:cnn_block_arch}, all the operations can be categorized into the following three types: depthwise convolution $3\times 3$ (including strides=1 and 2), pointwise convolution $1\times 1$ (including strides=1 and 2), and depthwise deconvolution $3 \times 3$. These three operations are implemented into three separated computing blocks. Input feature maps are fed to the corresponding block by dispatcher according to a pre-defined routine. The structure of computing engine is shown in Fig.~\ref{fig:pe_arch}.

\subsubsection{Pointhwise convolution}
Since pointweise convolution (or $1\times 1$ convolution) is literally vector matrix multiplication. 32 multiplier arrays (size=$32\times 1$) with corresponding 1 sum-32 adder tree and 1 accumulator for each array are implemented in PE. The adders in this block have higher precision than adders in other computing blocks.

\subsubsection{Depthwise convolution}
Depthwise convolution block utilizes the conventional architecture. 32 multiplier arrays (size=$3\times 3$) together with 32 sum-9 adder trees formed the depthwise convolution block.

\subsubsection{Depthwise deconvolution}
The implementation of depthwise deconvolution unit reuses the structure described in Sec.~\ref{sec:opt_dc}

The differences between deconvolution block and convolution block is 1) the patch of input feature map is $2\times 2$ instead of $3\times 3$ and 2) output 4 elements sequentially instead of 1.

\subsubsection{Activation Functions}
The activation function used in this neural network is Leaky ReLU, whose mathematical expression is
$$O_{L\!e\!a\!k\!y\!R\!e\!L\!U}=0.2\times min(x,0)+max(x,0)$$

This design can be easily extend to support other activation functions similar to this one, like ReLU, ReLU6 and etc. 

\subsection{Memory Mapping}
\subsubsection{Buffer for parameters}
Benefiting from the depthwise separable concept, the total number of parameters are reduced to less than 150K. This makes the parameters loading time reasonable low. Considering the parameter size for on-chip memory is still large, to balance the loading time and resource consumption, on-chip memory for half of the parameters are assigned.

\subsubsection{Buffer for feature maps}
Due to the relative large size of the feature maps and limitation of available on-chip memory, efficient mapping to reduce data communication is necessary. Partial unrolling results in loading same feature maps multiple times, which consequently requires higher memory bandwidth. To alleviate the burden, 10 feature map buffers with size $152\times 32\times 32$ are mapped. So that for layers whose channel is less than 256, no data transmitting is needed.

\begin{figure}[hbpt]
    \centering
    \includegraphics[width=0.9\columnwidth]{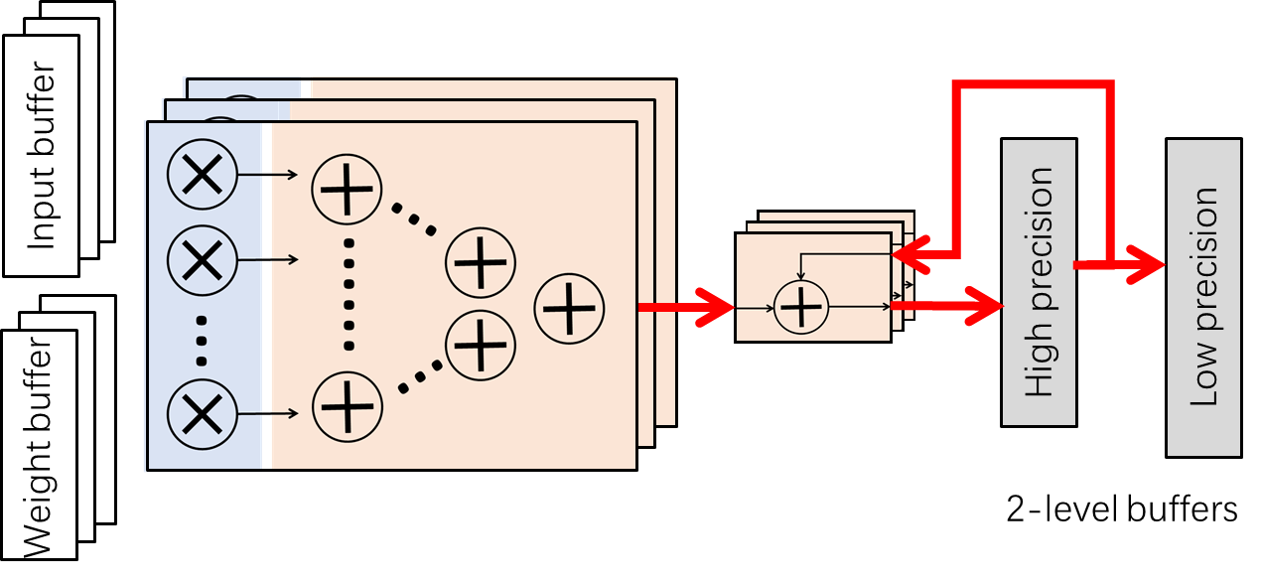}
    \caption{Extra buffer for pointwise convolution}
    \label{fig:mem_hier}
\end{figure}

Besides, one extra feature map buffers with same size but higher precision (or longer bitwidth) is mapped also. They are used for pointwise convolution only, aiming to decrease the precision loss (Fig.~\ref{fig:mem_hier}).

\section{Results and Discussion}\label{sec:result}
\begin{figure}[hbpt]
    \centering
    \includegraphics[width=\columnwidth]{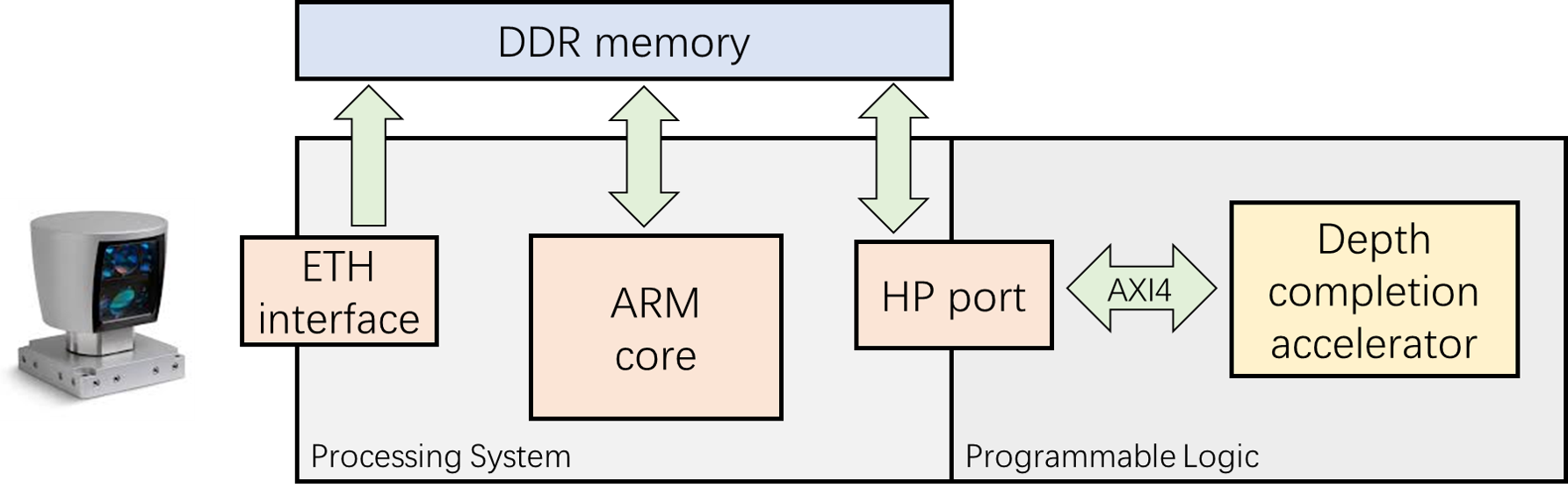}
    \caption{The overview of test setup}
    \label{fig:test_setup}
\end{figure}

This system has been implemented based on PYNQ open source framework \cite{xilinx_pynq} running on Xilinx ZCU104 Development Kit. The test setup is demonstrated in Fig.~\ref{fig:test_setup}, where LiDAR is connected to ZCU104 board via Ethernet cable. Velodyne LiDAR driver has been modified and loaded into Linux OS as dynamically linked shared object libraries. Users can send Python commands to the ARM processor who receives point cloud from LiDAR and stores it into DDR after pre-processing.

\begin{figure}[hbpt]
    \centering
    \includegraphics[width=0.9\columnwidth]{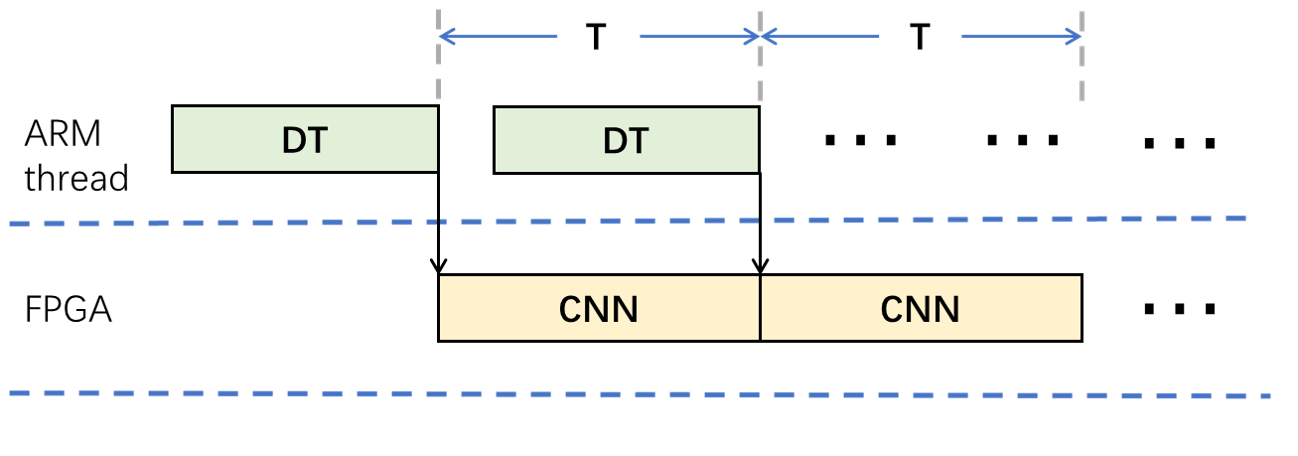}
    \caption{Hardware and software partitioning in time series}
    \label{fig:swhw_part}
\end{figure}

Before sending images into FPGA side, the ARM processor also does DT on the input point cloud. The software and hardware partitioning is illustrated in Fig.~\ref{fig:swhw_part}. When running at 200 MHz, this CNN accelerator can process 1 frame of point cloud within \textcolor{black}{90.1}ms. The total number of operations in this CNN inference is \textcolor{black}{15.14G}. Therefore, this accelerator achieves the computational capacity at \textcolor{black}{168.1}GOPS.

The hardware resource consumption is summarized in Tab.~\ref{tab:resource}. The bottleneck of this design is DSP resources, \textcolor{black}{98.1}\% of whom are mapped. Increasing the parallelism results in a large number of extra DSP slice utilisation. Besides, due to the large feature map size, around 87.5\% on-chip memory,including both BRAM and URAM, are utilized to buffer as much feature maps or parameters as possible.

\begin{table}[htb]
\centering
\caption{Resource consumption of depth completion CNN}
    \begin{tabular}{|c|c|c|c|c|c|} 
        \hline
         name & FF & LUT  & DSP  & BRAM & URAM \\
        \hline
        \hline
        FM buffer & 0 & 0 & 0 & 0 & 84 \\
        \hline
        weight buffer & 0 & 0 & 0 & 384 & 0 \\
        \hline
        dw conv & 49559 & 55209 & 288 & 64 & 0 \\
        \hline
        dw deconv & 49558 & 55208 & 288 & 64 & 0 \\
        \hline
        pw conv & 24624 & 36182 & 1057 & 2 & 0 \\
        \hline
        others & 3611 & 5261 & 64 & 31 & 0 \\
        \hline
        \hline
        \multirow{2}{2em}{Total} & 127352 & 151860 & 1695                    & 545                    & 84\\
                                 & (27.6\%) & (65.9\%) & (\textcolor{black}{98.1\%}) & (\textcolor{black}{87.3\%}) & (\textcolor{black}{87.5\%}) \\
        \hline
    \end{tabular}
\label{tab:resource}
\end{table}

\section{Conclusion}\label{sec:conclusion}

In this paper, we first propose a light-weight CNN namely DepthNet for the task of LiDAR point cloud depth completion. When comparing to state-of-the-art networks, DepthNet achieves similar error performance but only uses 3.8\% of parameters. Targeted for low-power embedded platforms such as autonomous vehicles, we further optimize the network with depthwise separable technique, which reduces the number of parameters by another factor of 7.3 at the cost of small degradation in error performance. Furthermore, we develop an FPGA system-on-chip that receives LiDAR data as input and produces dense depth maps. When evaluating with Velodyne HDL-64E LiDAR, it successfully demonstrates efficient and precise LiDAR depth completion at 11.1 fps that meets the real-time requirement for autonomous vehicles.    

\bibliographystyle{IEEEtran}
\bibliography{IEEEabrv,depthCompFPGA.bib}

\begin{thebibliography}{10}
\providecommand{\url}[1]{#1}
\csname url@samestyle\endcsname
\providecommand{\newblock}{\relax}
\providecommand{\bibinfo}[2]{#2}
\providecommand{\BIBentrySTDinterwordspacing}{\spaceskip=0pt\relax}
\providecommand{\BIBentryALTinterwordstretchfactor}{4}
\providecommand{\BIBentryALTinterwordspacing}{\spaceskip=\fontdimen2\font plus
\BIBentryALTinterwordstretchfactor\fontdimen3\font minus
  \fontdimen4\font\relax}
\providecommand{\BIBforeignlanguage}[2]{{%
\expandafter\ifx\csname l@#1\endcsname\relax
\typeout{** WARNING: IEEEtran.bst: No hyphenation pattern has been}%
\typeout{** loaded for the language `#1'. Using the pattern for}%
\typeout{** the default language instead.}%
\else
\language=\csname l@#1\endcsname
\fi
#2}}
\providecommand{\BIBdecl}{\relax}
\BIBdecl

\bibitem{ma2019self}
F.~Ma, G.~V. Cavalheiro, and S.~Karaman, ``Self-supervised sparse-to-dense:
  self-supervised depth completion from lidar and monocular camera,'' in
  \emph{2019 International Conference on Robotics and Automation (ICRA)}.\hskip
  1em plus 0.5em minus 0.4em\relax IEEE, 2019, pp. 3288--3295.

\bibitem{ku2018defense}
J.~Ku, A.~Harakeh, and S.~L. Waslander, ``In defense of classical image
  processing: Fast depth completion on the cpu,'' in \emph{2018 15th Conference
  on Computer and Robot Vision (CRV)}.\hskip 1em plus 0.5em minus 0.4em\relax
  IEEE, 2018, pp. 16--22.

\bibitem{chodosh2018deep}
N.~Chodosh, C.~Wang, and S.~Lucey, ``Deep convolutional compressed sensing for
  lidar depth completion,'' in \emph{Asian Conference on Computer
  Vision}.\hskip 1em plus 0.5em minus 0.4em\relax Springer, 2018, pp. 499--513.

\bibitem{eldesokey2018propagating}
A.~Eldesokey, M.~Felsberg, and F.~S. Khan, ``Propagating confidences through
  cnns for sparse data regression,'' \emph{arXiv preprint arXiv:1805.11913},
  2018.

\bibitem{zhang2015optimizing}
C.~Zhang, P.~Li, G.~Sun, Y.~Guan, B.~Xiao, and J.~Cong, ``Optimizing fpga-based
  accelerator design for deep convolutional neural networks,'' in
  \emph{Proceedings of the 2015 ACM/SIGDA International Symposium on
  Field-Programmable Gate Arrays}, 2015, pp. 161--170.

\bibitem{liu2018optimizing}
S.~Liu, H.~Fan, X.~Niu, H.-c. Ng, Y.~Chu, and W.~Luk, ``Optimizing cnn-based
  segmentation with deeply customized convolutional and deconvolutional
  architectures on fpga,'' \emph{ACM Transactions on Reconfigurable Technology
  and Systems (TRETS)}, vol.~11, no.~3, pp. 1--22, 2018.

\bibitem{zhang2017design}
X.~Zhang, S.~Das, O.~Neopane, and K.~Kreutz-Delgado, ``A design methodology for
  efficient implementation of deconvolutional neural networks on an fpga,''
  \emph{arXiv preprint arXiv:1705.02583}, 2017.

\bibitem{lyu2018real}
Y.~Lyu, L.~Bai, and X.~Huang, ``Real-time road segmentation using lidar data
  processing on an fpga,'' in \emph{2018 IEEE International Symposium on
  Circuits and Systems (ISCAS)}.\hskip 1em plus 0.5em minus 0.4em\relax IEEE,
  2018, pp. 1--5.

\bibitem{peng2019multi}
J.~Peng, L.~Tian, X.~Jia, H.~Guo, Y.~Xu, D.~Xie, H.~Luo, Y.~Shan, and Y.~Wang,
  ``Multi-task adas system on fpga,'' in \emph{2019 IEEE International
  Conference on Artificial Intelligence Circuits and Systems (AICAS)}.\hskip
  1em plus 0.5em minus 0.4em\relax IEEE, 2019, pp. 171--174.

\bibitem{he2016deep}
K.~He, X.~Zhang, S.~Ren, and J.~Sun, ``Deep residual learning for image
  recognition,'' in \emph{Proceedings of the IEEE conference on computer vision
  and pattern recognition}, 2016, pp. 770--778.

\bibitem{uhrig2017sparsity}
J.~Uhrig, N.~Schneider, L.~Schneider, U.~Franke, T.~Brox, and A.~Geiger,
  ``Sparsity invariant cnns,'' in \emph{2017 International Conference on 3D
  Vision (3DV)}.\hskip 1em plus 0.5em minus 0.4em\relax IEEE, 2017, pp. 11--20.

\bibitem{sandler2018mobilenetv2}
M.~Sandler, A.~Howard, M.~Zhu, A.~Zhmoginov, and L.-C. Chen, ``Mobilenetv2:
  Inverted residuals and linear bottlenecks,'' in \emph{Proceedings of the IEEE
  conference on computer vision and pattern recognition}, 2018, pp. 4510--4520.

\bibitem{van2019sparse}
W.~Van~Gansbeke, D.~Neven, B.~De~Brabandere, and L.~Van~Gool, ``Sparse and
  noisy lidar completion with rgb guidance and uncertainty,'' in \emph{2019
  16th International Conference on Machine Vision Applications (MVA)}.\hskip
  1em plus 0.5em minus 0.4em\relax IEEE, 2019, pp. 1--6.

\bibitem{tang2019learning}
J.~Tang, F.-P. Tian, W.~Feng, J.~Li, and P.~Tan, ``Learning guided
  convolutional network for depth completion,'' \emph{arXiv preprint
  arXiv:1908.01238}, 2019.

\bibitem{ma2017optimizing}
Y.~Ma, Y.~Cao, S.~Vrudhula, and J.-s. Seo, ``Optimizing loop operation and
  dataflow in fpga acceleration of deep convolutional neural networks,'' in
  \emph{Proceedings of the 2017 ACM/SIGDA International Symposium on
  Field-Programmable Gate Arrays}, 2017, pp. 45--54.

\bibitem{bai_iscas2020}
L.~Bai, Y.~Lyu, and X.~Huang, ``A unified hardware architecture for
  convolutions and deconvolutions in cnn,'' in \emph{2018 IEEE International
  Symposium on Circuits and Systems (ISCAS)}.\hskip 1em plus 0.5em minus
  0.4em\relax IEEE, 2020, pp. 1--5.

\bibitem{xilinx_pynq}
\BIBentryALTinterwordspacing
{Xilinx Inc.}, ``pynq.io,'' 2016. [Online]. Available: \url{www.pynq.io}
\BIBentrySTDinterwordspacing

\end{thebibliography}

\end{document}